# Rational Design of Two-Dimensional Octuple-Atomic-Layer $M_2A_2Z_4$ for Photocatalytic Water Splitting

Shikai Chang[a], Dingyanyan Zhou[a], Yujin Ji*[a], Mir F. Mousavi[b], Jian Xi*[c], and Youyong Li*[a,d]

[a] S. Chang, D. Zhou, Y. Ji, Y. Li
State Key Laboratory of Bioinspired Interfacial Materials Science
Institute of Functional Nano & Soft Materials (FUNSOM)
Soochow University
Suzhou 215123, China
E-mail: yjji@suda.edu.cn (Y. Ji)

[b] M. F. Mousavi
Department of Chemistry
Faculty of Basic Sciences
Tarbiat Modares University
Tehran

[c] J. Xi
Suzhou Laboratory
Suzhou 215123, China
E-mail: xij@szlab.ac.cn

[d] Y. Li
Macao Institute of Materials Science and Engineering
Macau University of Science and Technology
Taipa, Macau SAR 999078, China
E-mail: yyli@suda.edu.cn (Y. Li)

Supporting information for this article is given via a link at the end of the document.

**Abstract:** Two-dimensional (2D) materials have emerged as promising candidates as photocatalytic materials due to their large surface areas and tunable electronic properties. In this work, we systematically design and screen a series of octuple-atomic-layer $M_2A_2Z_4$ monolayers (M = Al, Ga, In; A = Si, Ge, Sn; Z = N, P, As) using first-principles calculations. 108 structures are constructed by intercalation approach, followed by a comprehensive evaluation of their thermodynamic and dynamic stability, band gaps, and band edge alignments to assess their potential for photocatalytic overall water splitting. Eight candidates meet the criteria for overall water splitting, among which $Al_2Si_2N_4$ and $Al_2Ge_2N_4$ exhibit suitable band edge positions, pronounced visible-light absorption, high electron mobility and high solar-to-hydrogen (STH) efficiencies for photocatalysis under both acidic and neutral environments (pH = 0 and 7). Importantly, the introduction of N vacancies on the surfaces of $Al_2Si_2N_4$ and $Al_2Ge_2N_4$ significantly enhances their catalytic activity for both hydrogen reduction and water oxidation reactions, further supporting their potential as photocatalysts for overall water splitting. Both materials also display robust structural stability in aqueous environments. Our study provides theoretical insights for the rational design of efficient and stable 2D photocatalysts for overall water splitting.

## Introduction

Photocatalytic water splitting has emerged as a promising solar-to-hydrogen energy conversion strategy, driven by urgent needs to address fossil fuel depletion and environmental challenges.[1–5] The efficiency of this technology is fundamentally governed by the development of high-performance photocatalysts. An ideal photocatalyst typically requires: 1. Robust stability under operational conditions; 2. An optimal band gap for board solar spectrum utilization; 3. Well-aligned band edge positions to thermodynamically drive redox reactions; 4. Efficient charge carrier separation and migration.[6–9] These stringent criteria underscore the importance of systematic material design in advancing this field.

Atomic-layer-dependent two-dimensional (2D) materials can offer both large surface areas for photochemical reactions and tunable electronic structures for charge transfer, making them desirable for efficient photocatalysis.[10–12] For instance, g-$C_3N_4$[13–15] and phosphorene[16,17] serve as representative single-atomic-layer and double-atomic-layer 2D photocatalysts. Significant efforts have been dedicated to triple-atomic-layer transition metal dichalcogenides (TMDs),[18–22] whose optimal bandgap alignment and high carrier mobility enable efficient photocatalytic processes. Other triple-atomic-layer $Pt_5Se_4$,[23] quadruple-atomic-layer Group-III chalcogenides, e.g., GaSe[24] and InSe,[25,26] and quintuple-atomic layer materials exemplified by $Bi_2Se_3$[27] further enrich the family of 2D photocatalysts by offering tunable electronic properties. Recent breakthrough in synthesizing $MoSi_2N_4$ lead to the discovery of the septuple-atomic-layer $MA_2Z_4$ family, among which $MoSi_2N_4$, $WSi_2N_4$, and $WGe_2N_4$ are revealed as attractive candidates for photocatalytic water splitting.[28] With the increase in atomic-layer thickness, 2D materials exhibit greater compositional diversity and structural tunability, thereby offering enhanced opportunities for optimizing photocatalytic performance.

The intercalation approach for constructing 2D $MA_2Z_4$ structures has provided a feasible strategy for designing 2D materials with multiple atomic layer thicknesses.[29,30] In this strategy, a $MoS_2$-like $MZ_2$ is intercalated into an InSe-type $A_2Z_2$, forming a stable layered $MA_2Z_4$ structure. This intercalation design approach not only stabilizes the system but also modifies the band structures of the corresponding 2D components, resulting in enriched electronic properties. To date, research on $MA_2Z_4$ materials primarily focused on the structures with transition metals (e.g., Mo and W) as the central atoms, which can be readily extended to other elemental groups, such as IIIA metals (e.g., Al, Ga, and In).[31] In particular, IIIA nitrides, as third-generation semiconductors, including AlN, GaN, and InN, possess tunable direct band gaps, high carrier mobility, and excellent chemical stability, making them promising for applications in optical devices.[32–34] Theoretical results have shown that 2D AlN, GaN, and InN monolayers adopt graphene-like structures with significantly larger band gaps compared with their 3D bulk materials and high exciton binding energies ranging from 0.6 to 1.9 eV.[35]

Following this idea, the surface of IIIA-based $M_2Z_2$ can be passivated with the $A_2Z_2$ layers to construct the $M_2A_2Z_4$ materials with modulated band-edge positions and distinct electronic properties. In this study, we systematically design a series of 2D octuple-atomic-layer $M_2A_2Z_4$ (M = Al, Ga, In; A = Si, Ge, Sn; Z = N, P, As) monolayers and investigate their structural, electronic and photocatalytic properties through first-principles calculations. We first construct and optimize 108 $M_2A_2Z_4$ configurations and evaluate their structural stability. Subsequently, we assess their electronic structures, including band gaps and band edge alignments. Our results reveal that $Al_2Si_2N_4$ and $Al_2Ge_2N_4$ are promising candidates under both acidic and neutral conditions. By further examining their optical absorption, carrier mobility, solar-to-hydrogen (STH) efficiency, catalytic activity for both the hydrogen evolution reaction (HER) and the oxygen evolution reaction (OER), and stability in aqueous environments, we confirm their potential for efficient overall water splitting. This work enriches the $M_2A_2Z_4$ material family and provides insights for the design of efficient 2D photocatalysts.

## Results and Discussion

As illustrated in Figure 1a, we construct hexagonal $M_2A_2Z_4$ monolayers by intercalating an $M_2Z_2$ layer into a $A_2Z_2$ layer following the intercalation strategy. Here, M, A, and Z denote IIIA metals (M = Al, Ga, In), IVA elements (A = Si, Ge, Sn), and VA elements (Z = N, P, As), respectively. During the construction, two types of phases of *α* and *β* phases are considered for each constituent. The *α* phase crystallizes in the $P\bar{6}m2$ space group, and the *β* phase belongs to $P\bar{3}m1$. Based on symmetry, four types of combinations of $M_2Z_2$ and $A_2Z_2$ are generated, namely *αα*, *αβ*, *βα*, and *ββ*, where the first and second symbols denote the phases of $M_2Z_2$ and $A_2Z_2$, respectively. In total, 108 candidate $M_2A_2Z_4$ structures are constructed for subsequent investigation. The corresponding structure files are provided in the Supporting Information.

Ensuring the stability of $M_2A_2Z_4$ is a prerequisite for further property evaluation. We first calculate the formation energies to assess their thermodynamic stability. Figure 1b shows the formation energies of four phase for each $M_2A_2Z_4$ compound, referenced to the total energy of the *αα* phase. The results indicate that for $M_2Si_2N_4$ (M = Al, Ga, In), the *αα* phase is the most stable. For $M_2Sn_2N_4$ (M = Al, Ga, In), the *αβ* phase is identified as the most energetically favorable. In the cases of $M_2Sn_2P_4$ and $M_2A_2As_4$ (M = Al, Ga, In; A = Si, Ge, Sn), the *βα* phase shows the lowest energy, except for $Ga_2Si_2As_4$ and $In_2Sn_2As_4$. Moreover, $M_2A_2Z_4$ materials containing Ge or In tend to exhibit more diverse thermodynamically stable phases, and information of the formation energies of $M_2A_2Z_4$ are detailed in Table S1. To further confirm their structural stability, phonon dispersion analyses are performed, as shown in Figures S1-S3. Our results suggest that 26 out of the 27 thermodynamically favorable $M_2A_2Z_4$ are dynamically stable. Only $In_2Sn_2P_4$ shows imaginary phonon branches, indicating dynamical instability. These stability analyses can demonstrate the effectiveness of our intercalation-based approach for constructing stable $M_2A_2Z_4$ monolayers.

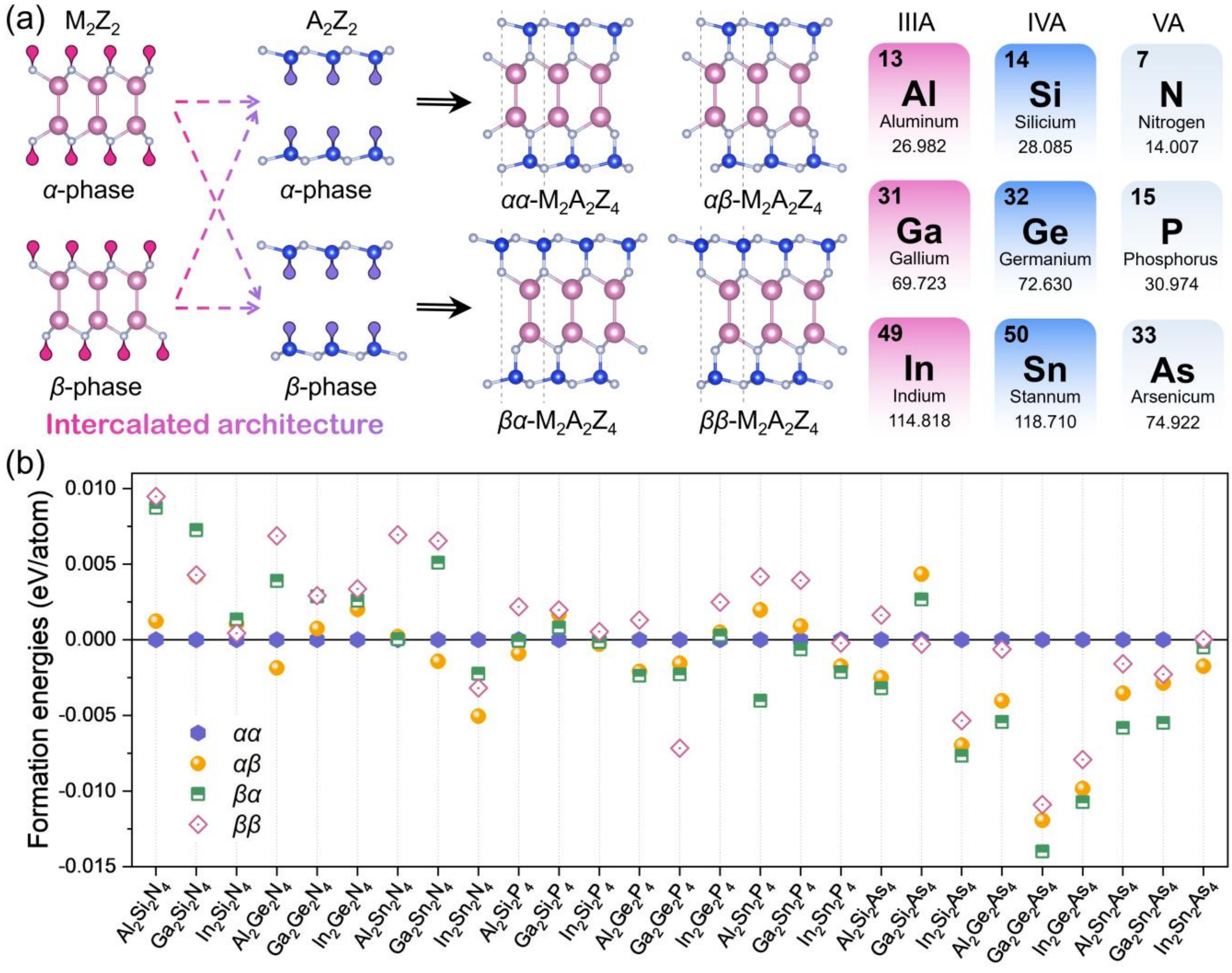

**Figure 1**. (a) Schematic illustration of the construction method, four possible structures, and elemental compositions of $M_2A_2Z_4$. Pink, blue, and silver spheres represent M, A, and Z atoms, respectively. (b) Comparison of formation energies among four structural configurations of $M_2A_2Z_4$ monolayers. The formation energy of each compound is relative to the energy of the *αα* structure.

Stability assessment forms the basis of our screening, after which electronic structures and photocatalytic potential are examined. The overall screening process is displayed in Figure 2a. We calculate their electronic band structures of 26 $M_2A_2Z_4$ candidates, as shown in Figures S4-S6. Both PBE and hybrid HSE functional calculations are applied for the band structures to achieve the accurate results. Detailed band gap values are summarized in Table S2. Our results demonstrate that $M_2A_2Z_4$ show sizeable band gaps varying from 0.26 eV ($In_2Ge_2As_4$) to 3.69 eV ($Ga_2Si_2N_4$). When Z = N or P, the corresponding monolayers tend to exhibit larger band gaps than those with Z = As. As an example, $Al_2Ge_2N_4$ has an indirect band gap of 1.76 eV, slightly higher than that of $Al_2Ge_2P_4$ (1.66 eV), while $Al_2Ge_2As_4$ shows a further reduced gap of 1.19 eV. This trend can be attributed to the stronger electronegativity and smaller atomic radius of the N or P atom, which enhance orbital overlap and consequently widen the energy separation between the valence and conduction bands. Similarly, for M atoms, monolayers containing Al or Ga exhibit larger band gaps than those with In. For example, $Ga_2Si_2N_4$ displays a band gap of 3.69 eV, significantly larger than that of $In_2Si_2N_4$, which is 2.54 eV. It is notable that among the structural stable $M_2A_2Z_4$ compounds studied, most exhibit indirect band gaps, and only $In_2Si_2P_4$, $In_2Ge_2P_4$, $In_2Ge_2P_4$, $In_2Si_2As_4$, and $In_2Ge_2As_4$ possess direct band gaps. For photocatalytic water splitting, an ideal photocatalyst must possess a band gap larger than 1.23 eV to thermodynamically drive the overall reaction. Meanwhile, to ensure efficient solar absorption, the band gap should also remain below an upper threshold of approximately 2.50 eV. Based on this criterion (1.23 eV < $E_g$ < 2.50 eV), 15 semiconducting $M_2A_2Z_4$ materials are identified as potential candidates for photocatalytic water splitting due to their suitable band gaps. To further assess the feasibility of $M_2A_2Z_4$ monolayers as overall water splitting photocatalysts, we further align their band edge positions with vacuum level corrections and compared with the potentials of hydrogen reduction (−4.44 eV) and water oxidation (−5.67 eV) reactions. As shown in Figure 2b, under acidic conditions (pH = 0), the conduction band minimum (CBM) positions of $Al_2Si_2N_4$, $Ga_2Si_2N_4$, $Al_2Si_2P_4$, $Ga_2Si_2P_4$, $Al_2Ge_2P_4$, $Ga_2Ge_2P_4$, $Al_2Sn_2P_4$, and $Al_2Si_2As_4$ are higher than the hydrogen reduction potential, while their valence band maximum (VBM) positions are lower than the water oxidation potential, thus fulfilling the thermodynamic requirements for overall water splitting. Moreover, at pH = 0, the VBM positions of 12 candidate $M_2A_2Z_4$ materials, excluding $Ga_2Si_2As_4$, $Al_2Sn_2As_4$, and $Ga_2Sn_2As_4$, are lower the water oxidation potential, indicating their capability to facilitate the water oxidation reaction. Likewise, the CBM positions of most compounds, except for $In_2Ge_2N_4$, $In_2Sn_2N_4$, $In_2Si_2P_4$, and $Ga_2Sn_2P_4$, are higher than the hydrogen reduction potential, suggesting their ability to support the hydrogen reduction reaction. Notably, under neutral conditions (pH = 7), $Al_2Si_2N_4$ and $Al_2Ge_2N_4$ maintain CBM positions above the hydrogen reduction potential (−4.03 eV), indicating their promising ability to drive overall water splitting across both acidic and neutral environments.

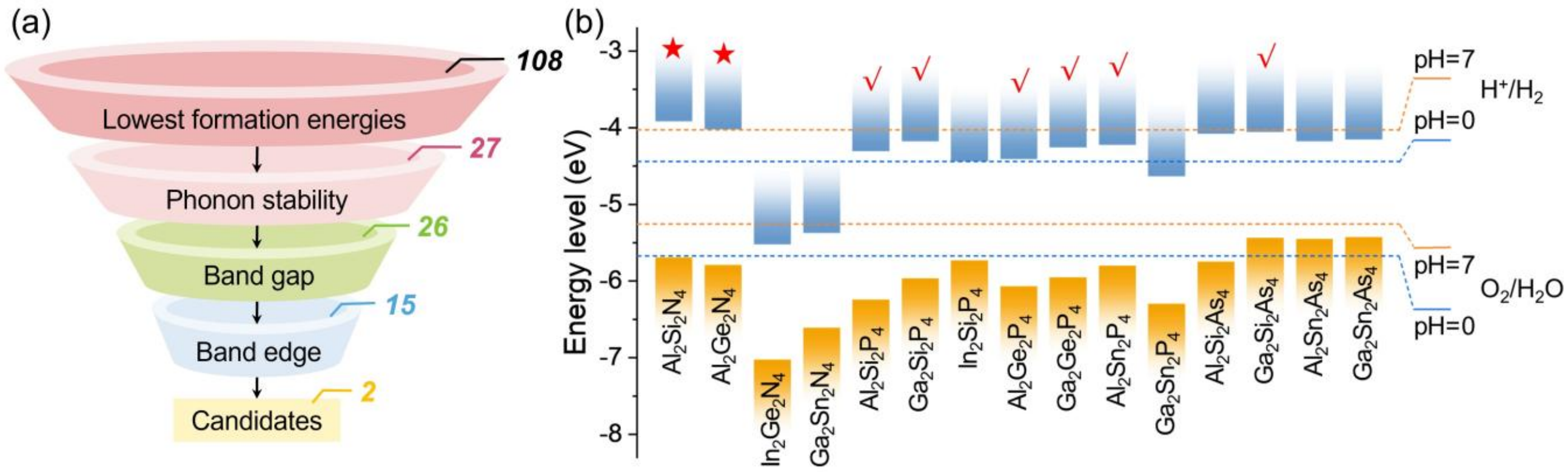

**Figure 2**. (a) Strategy for screening promising $M_2A_2Z_4$ monolayers as photocatalysts for overall water splitting. (b) The band edge alignments of $M_2A_2Z_4$ materials based on HSE06 calculations with respect to standard water redox potentials.

To gain deeper insights into the photocatalytic potential of $M_2A_2Z_4$ materials, we conduct a comprehensive investigation of two promising candidates $Al_2Si_2N_4$ and $Al_2Ge_2N_4$. An essential requirement for efficient photocatalysts is their ability to effectively harvest solar energy, particularly in the ultraviolet and visible light regions. To evaluate this property, we obtain the optical absorption coefficients of $Al_2Si_2N_4$ and $Al_2Ge_2N_4$ based the HSE06 hybrid functional calculations, as presented in Figure 3. The results reveal that both $Al_2Si_2N_4$ and $Al_2Ge_2N_4$ exhibit pronounced optical absorption in the visible regions of the solar spectrum, specifically between 2.0 and 2.3 eV (corresponding to wavelengths of approximately 540–620 nm), indicating their strong responses to visible light. $Al_2Si_2N_4$ also shows additional absorption features at 3.0 eV (visible region, ~413 nm) and 3.7 eV (ultraviolet region, ~335 nm). Meanwhile, $Al_2Ge_2N_4$ presents distinct optical absorption at 2.8 eV (~443 nm), 3.4 eV (~365 nm), and 3.8 eV (~326 nm). These results suggest that both materials possess ideal optical absorption across a broad spectral range spanning from visible to ultraviolet light.

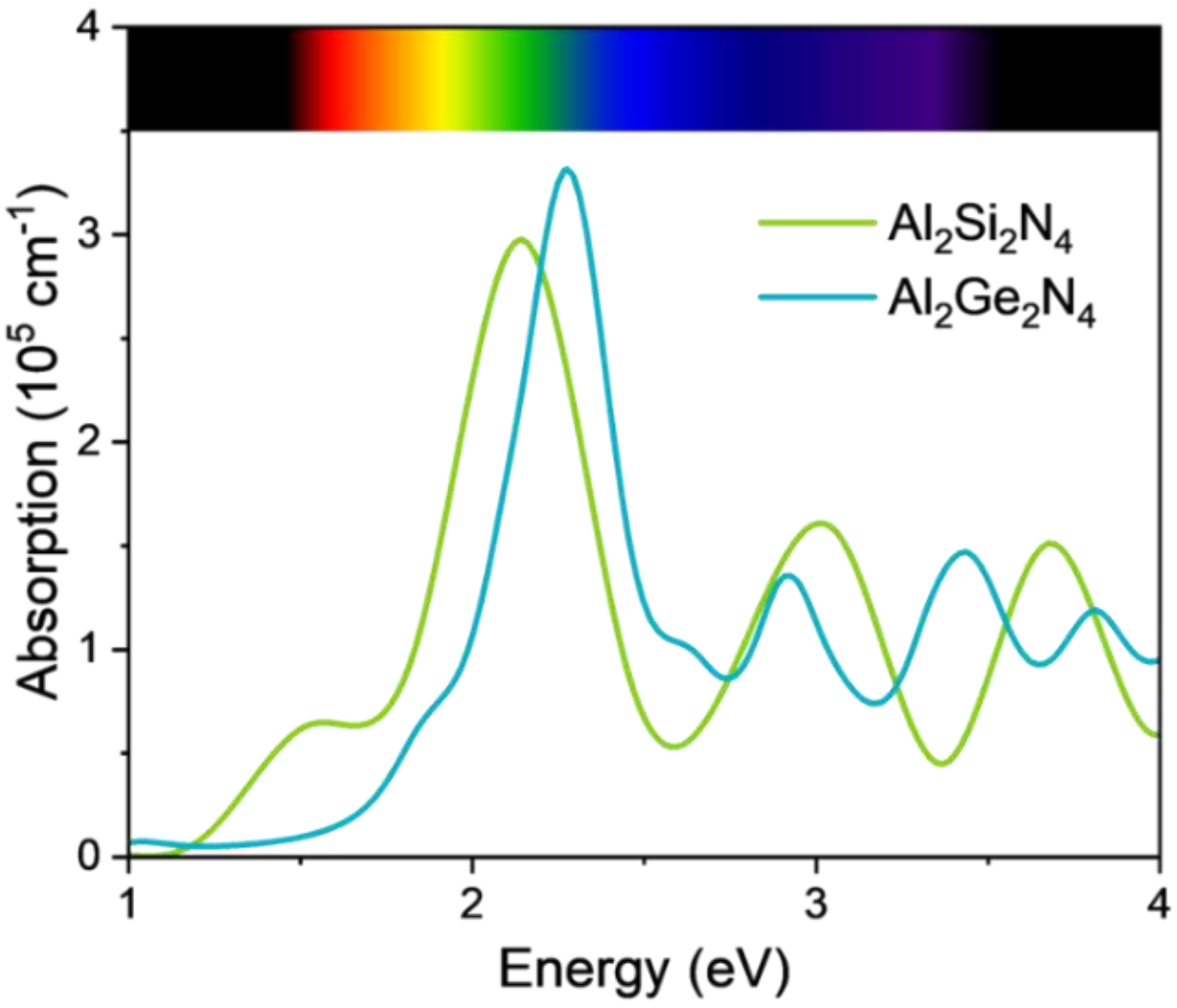

**Figure 3**. Optical absorption spectra of $Al_2Si_2N_4$ and $Al_2Ge_2N_4$ using the HSE06 functional.

Then carrier mobility is also an important indicator of a good photocatalyst, we calculate the carriers' effective masses and mobilities along x and y directions using orthorhombic lattices accordingly. The obtained effective masses ($m^*$), deformation potential constant ($E_1$), elastic modulus ($C_{2D}$), and carrier mobilities ($\mu$) are listed in Table 1. For $Al_2Si_2N_4$, the electron mobility along the x and y directions reaches 2504.04 and 8235.10 cm$^2$ V$^{-1}$ s$^{-1}$, respectively, while the corresponding hole mobility is 207.36 and 1643.41 cm$^2$ V$^{-1}$ s$^{-1}$, respectively. This anisotropic difference arises because the effective masses of holes is greater than that of electrons. $Al_2Ge_2N_4$ exhibits a similar trend. For $Al_2Ge_2N_4$, the electron mobility in the x and y directions is 1099.94 and 686.73 cm$^2$ V$^{-1}$ s$^{-1}$, respectively, while the hole mobility is 149.64 and 432.56 cm$^2$ V$^{-1}$ s$^{-1}$.

Table S3. Calculated the effective mass $m^*$, deformation potential constant $E_1$, elastic modulus $C_{2D}$, carrier mobility $\mu$ along the x and y direction at 300 K

| Material | | $m^*$ ($m_0$) | | $E_1$ (eV) | | $C_{2D}$ (N/m) | $\mu$ (cm$^2$ V$^{-1}$ s$^{-1}$) | |
|---|---|---|---|---|---|---|---|---|
| | | e | h | e | h | e = h | e | h |
| $Al_2Si_2N_4$ | x | 0.88 | 1.18 | 3.33 | 8.52 | 563.82 | 2504.04 | 207.36 |
| | y | 0.27 | 0.38 | 3.33 | 5.36 | 568.98 | 8235.10 | 1643.41 |
| $Al_2Ge_2N_4$ | x | 0.27 | 1.41 | 11.39 | 7.71 | 481.86 | 1099.94 | 149.64 |
| | y | 0.27 | 0.46 | 14.37 | 7.89 | 480.55 | 686.73 | 432.56 |

Moreover, in order to evaluate their potential applications of the selected $Al_2Si_2N_4$ and $Al_2Ge_2N_4$, their theoretical STH efficiencies are further assessed. The STH efficiency of a photocatalyst is intrinsically related to pH conditions.

Here, we employ the PySTH program[36] to evaluate the theoretical STH efficiencies of $M_2A_2Z_4$ systems under different pH conditions. As shown in Figure 4, pH regulation enables the enhancement for STH efficiencies of $Al_2Si_2N_4$ and $Al_2Ge_2N_4$, which increase from 10.07% and 12.21% at pH = 0 to 17.12% at pH = 6–7, respectively.

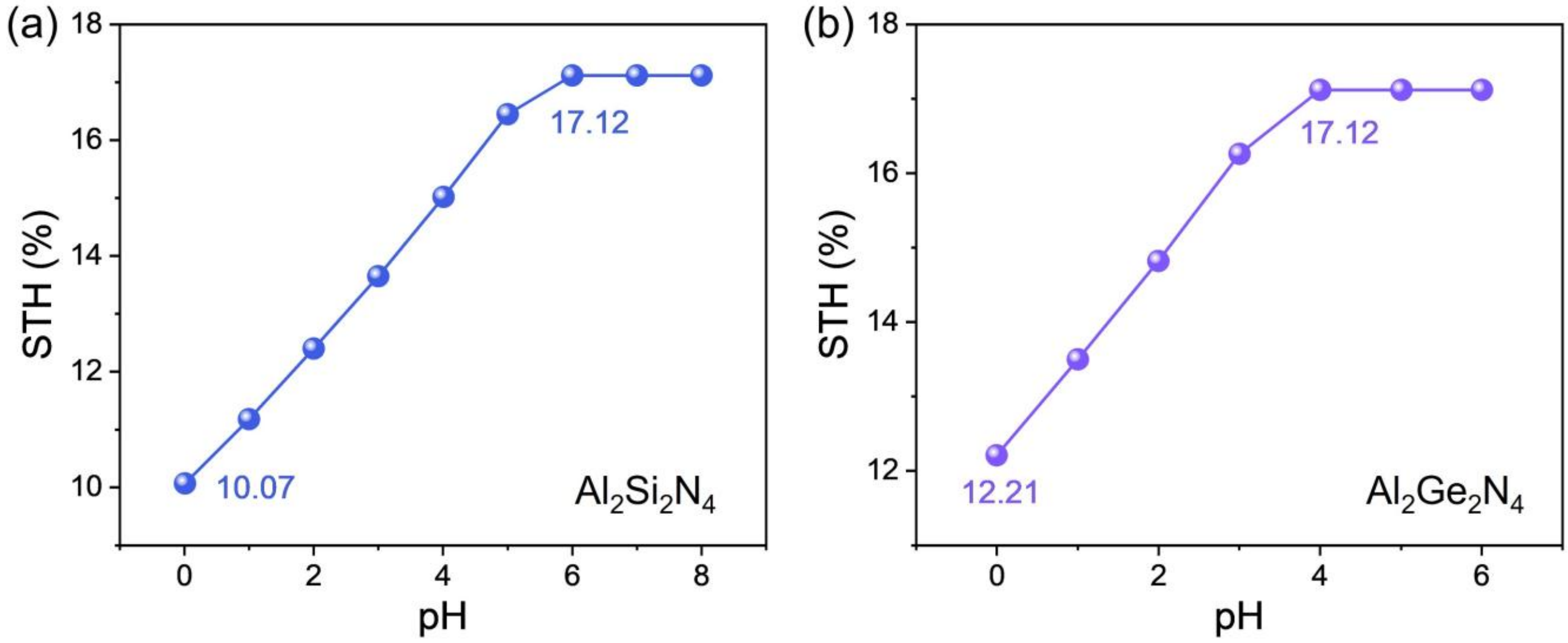

**Figure 4**. Variation of STH efficiency with pH for (a) $Al_2Si_2N_4$ and (b) $Al_2Ge_2N_4$.

To further investigate the overall water splitting photocatalytic proficiency of $Al_2Si_2N_4$ and $Al_2Ge_2N_4$, we analyze the thermodynamics of the two half-reactions of both hydrogen reduction reaction and water oxidation reaction. Figure 5 presents the HER performance of $Al_2Si_2N_4$ and $Al_2Ge_2N_4$. HER involves a two-electron reaction process, which is described as:

$$* + H^+ + e^- \rightarrow *H$$

$$*H + H^+ + e^- \rightarrow * + H_2$$

It is observed that on the surface of pristine $Al_2Si_2N_4$ and $Al_2Ge_2N_4$, hydrogen atoms are preferentially adsorbed on Si/Ge sites, with the HER Gibbs free energy ($\Delta G_H$) of 1.59 eV for $Al_2Si_2N_4$ and 1.25 eV for $Al_2Ge_2N_4$. However, the potential of photogenerated electrons for hydrogen reduction ($U_e$) of these two materials only provides a driving force of 0.52 V and 0.41 V, respectively, which is insufficient to overcome their HER energy barrier. This suggests that pristine $Al_2Si_2N_4$ and $Al_2Ge_2N_4$ have difficulty in spontaneously driving the HER, consistent with similar observations in $MoS_2$, $WSe_2$, and $MoSi_2N_4$.[37–40] However, some intrinsic vacancies formed during the synthesis of 2D materials can serve as active sites for reactions, and thereby regulate catalytic activity without significantly impairing structural stability and intrinsic properties.[40,41] Based on a comparative examination of the defect sites and their formation energies, as described in the Supporting Information, we focus on introducing N vacancies on the pristine surfaces of $Al_2Si_2N_4$ and $Al_2Ge_2N_4$. Upon introducing N vacancy, the hydrogen adsorption shifts from the initial Si/Ge site to the N vacancy. Consequently, the $\Delta G_H$ values are significantly reduced to 0.14 eV for $Al_2Si_2N_4$ and -0.32 eV for $Al_2Ge_2N_4$ (Figure 5), indicating that HER can proceed spontaneously under the driving force provided by $U_e$.

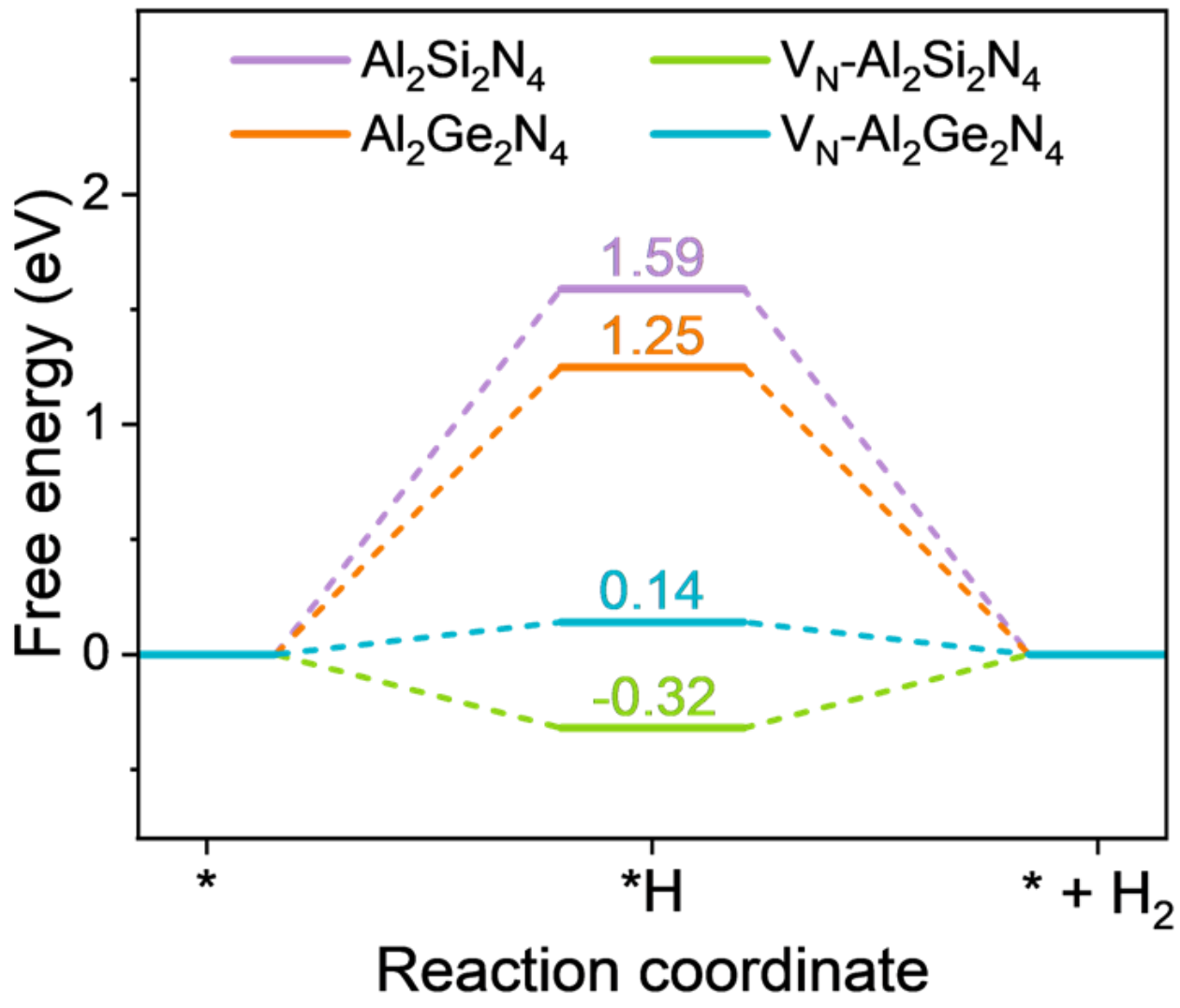

**Figure 5**. Gibbs free energy diagrams for hydrogen adsorption on pristine and N-vacancy defective $Al_2Si_2N_4$ and $Al_2Ge_2N_4$.

For OER half-reaction, we consider a four-electron step reaction, which involves the sequential formation of *OH, *O, and *OOH intermediates. Under pH = 0 condition, the overall reaction can be summarized as:

$$H_2O + * \rightarrow *OH + H^+ + e^-$$

$$*OH \rightarrow *O + H^+ + e^-$$

$$H_2O + *O \rightarrow *OOH + H^+ + e^-$$

$$*OOH \rightarrow O_2 + H^+ + e^-$$

We evaluate the Gibbs free energy differences (Δ*G*) of each OER step at pH = 0 for the pristine $Al_2Si_2N_4$ and $Al_2Ge_2N_4$, as illustrated in Figure S7. On the pristine surfaces of both materials, the rate-limiting step is identified as the transition of *OH → *OOH, with Δ*G* of 2.11 eV for $Al_2Si_2N_4$ and 2.06 eV for $Al_2Ge_2N_4$, which significantly exceed the potential of holes for water oxidation ($U_h$) they can provide (1.25 V and 1.36 V, respectively). However, as shown in Figure 6, with the introduction of N vacancy, the Δ*G* of the rate-limiting step decreases to 1.86 eV for $Al_2Si_2N_4$ and 1.76 eV for $Al_2Ge_2N_4$. Under light irradiation, the energies required for overcoming the rate-limiting step further drop to 0.30 eV and 0.28 eV, respectively. Moreover, when the pH value is adjusted to 7, the Gibbs free energies of all OER steps exhibit downhill for both $Al_2Si_2N_4$ and $Al_2Ge_2N_4$ materials, further supporting their potential as efficient OER photocatalysts.

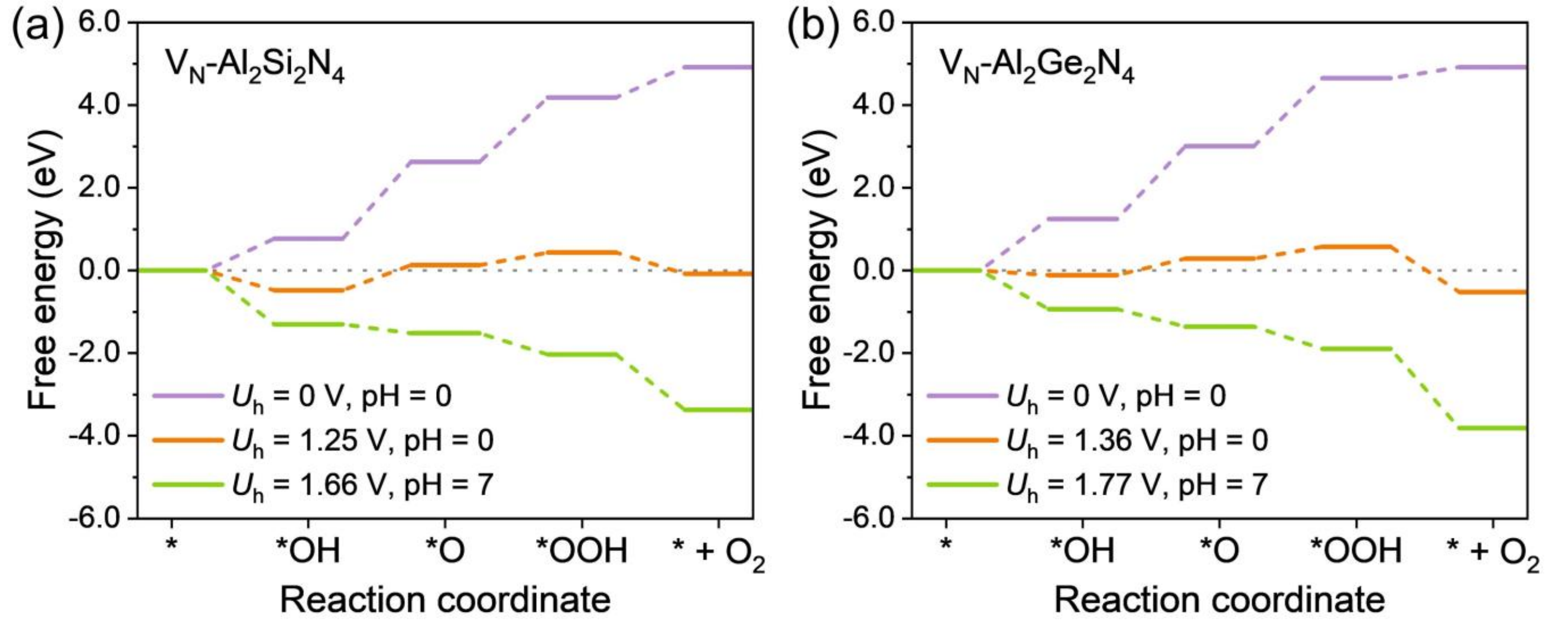

**Figure 6**. Gibbs free energy profiles of the OER on (a) $Al_2Si_2N_4$ and (b) $Al_2Ge_2N_4$ with N vacancy defects. The purple line represents conditions in dark at pH = 0, the orange line corresponds to conditions of light irradiation at pH = 0, and the green line represents conditions of light irradiation at pH = 7.

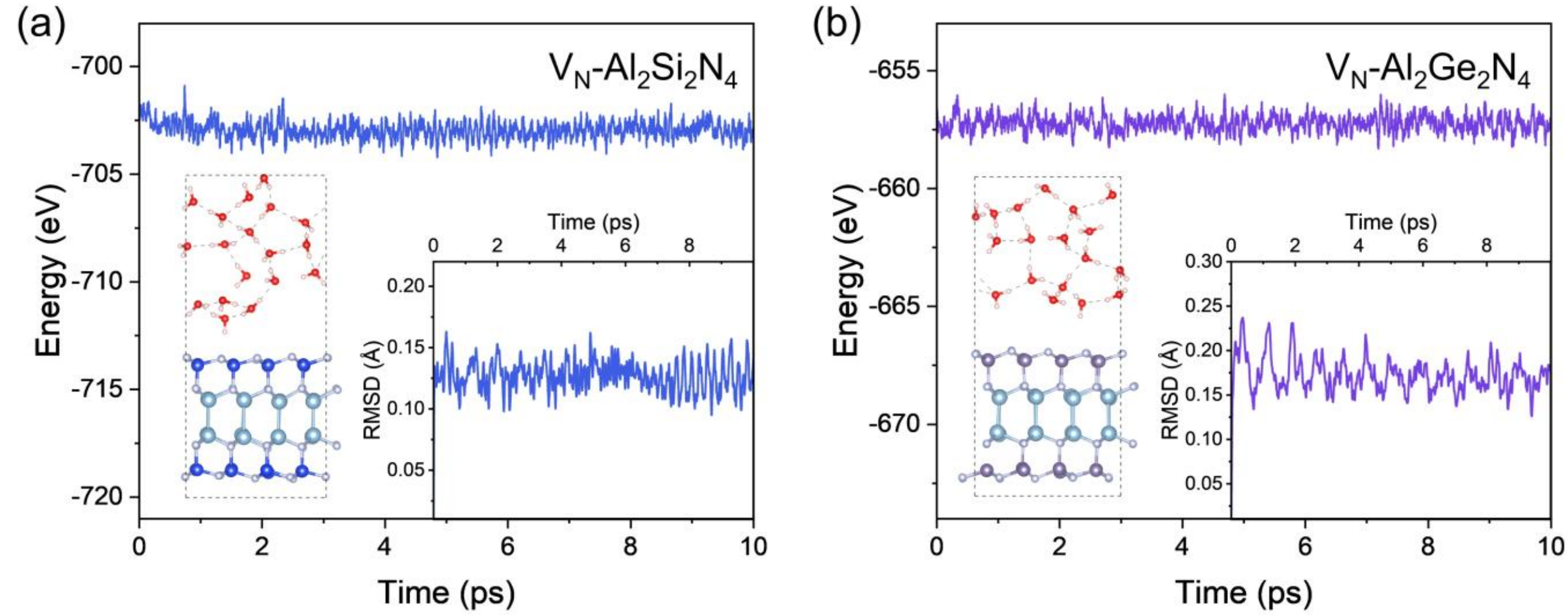

**Figure 7**. AIMD simulations and RMSD plots of (a) $Al_2Si_2N_4$ and (b) $Al_2Ge_2N_4$ with N vacancy defects under the explicit solvation effect.

Last but not least, we further assess the stability of $Al_2Si_2N_4$ and $Al_2Ge_2N_4$ with N vacancy defect in an aqueous environment. As shown in Figure 7 and Figure S8, Ab initio molecular dynamics (AIMD) simulations of $Al_2Si_2N_4$ and $Al_2Ge_2N_4$ with N vacancy defect are carried out for 10 ps at 300 K under explicit solvent effects, with water molecules placed on both sides of the non-periodic direction of the monolayers. Throughout 10 *ps* AIMD simulation, the root mean square deviation (RMSD) of $Al_2Si_2N_4$ fluctuate slightly around 0.12 Å, and that of $Al_2Ge_2N_4$ stabilized near 0.17 Å. These results suggest that both materials maintained their original 2D layered structures and bonding configurations without significant distortion or bond breakage. This demonstrates their structural stability under aqueous conditions, confirming their feasibility as photocatalysts for overall water splitting.

## Conclusion

To summarize, based on the intercalation strategy, we systematically design and screen 2D $M_2A_2Z_4$ materials as potential photocatalysts for overall water splitting based on first-principles calculations. 108 kinds of $M_2A_2Z_4$ structures are constructed, and through comprehensive evaluation of their thermodynamic and dynamic stability, band gaps, and band edge alignments, among which 8 candidates ($Al_2Si_2N_4$, $Ga_2Si_2N_4$, $Al_2Si_2P_4$, $Ga_2Si_2P_4$, $Al_2Ge_2P_4$, $Ga_2Ge_2P_4$, $Al_2Sn_2P_4$, and $Al_2Si_2As_4$) are identified as capable of driving overall water splitting at pH = 0. Notably, $Al_2Si_2N_4$ and $Al_2Ge_2N_4$ also possess the overall water splitting photocatalytic potential at pH = 7. The electron (hole) mobility of $Al_2Si_2N_4$ reaches values as high as 8235.10 (1643.41) $cm^2 V^{-1} s^{-1}$, while the maximum electron (hole) mobility of $Al_2Ge_2N_4$ is 1099.94 (432.56) $cm^2\,V^{-1}\,s^{-1}$. Both $Al_2Si_2N_4$ and $Al_2Ge_2N_4$ achieve an STH efficiency of 17.12 % under the neutral environment. Further investigations reveal that both materials demonstrate strong optical absorption in the visible spectrum region, enabling efficient utilization of solar energy. More importantly, when N vacancies are introduced on the surfaces of these two materials, they exhibit superior catalytic performance for both hydrogen reduction and oxygen oxidation reactions. At pH = 0, the presence of N vacancies optimizes the $\Delta G_H$ to 0.14 eV for $Al_2Si_2N_4$ and -0.32 eV for $Al_2Ge_2N_4$, suggesting their abilities to drive the HER effectively. when pH adjusted to 7, the Gibbs free energies of all OER steps decrease for both $Al_2Si_2N_4$ and $Al_2Ge_2N_4$ with N vacancies under light irradiation, indicating their viability as efficient OER photocatalysts. In addition, they can maintain excellent structural stability in aqueous environments, further supporting their potential as efficient photocatalysts for overall water splitting. Our findings offer insights for the design of stable and high-performance 2D photocatalysts.

## Computational Methods

All first-principles calculations are performed based on density functional theory (DFT) using the Vienna ab initio simulation package (VASP).[42] The Perdew–Burke–Ernzerhof (PBE) functional within the generalized gradient approximation (GGA) is used in the structural optimizations and electronic band structure calculations.[43,44] To obtain accurate band gaps and band edge alignments, the Heyd–Scuseria–Ernzerhof (HSE06) hybrid functional is employed.[45] The projector augmented-wave (PAW) method is used to describe the ion–electron interactions,[46] and the plane-wave energy cutoff is set to 500 eV. A vacuum space over 15 Å is applied along the non-periodic direction to eliminate interactions between periodic images. The Brillouin zone is sampled using a Γ-centered 12 × 12 × 1 k-point grid. During geometry optimizations, the convergence thresholds for energy and forces are set to $1\times10^{-6}$ eV and $1\times10^{-3}$ eV/Å, respectively. Phonon dispersions are calculated using the finite displacement method as implemented in the Phonopy package,[47,48] based on a 4 × 4 × 1 supercell. AIMD simulations are performed based on the Nosé−Hoover thermostat at the temperature of near 300 K.[49] The simulations last for 10 ps with a time step of 1 fs. The AIMD simulation cells are obtained by orthogonalizing the primitive cell and constructing a 2 × 2 × 1 supercell with Γ-centered 2 × 1 × 1 k-point sampling.

The formation energies of $M_2A_2Z_4$ monolayers are calculated as follows:

$$E_f = [E_{tot} - (2E_M + 2E_A + 4E_Z)] / 8 \quad (1)$$

where $E_{tot}$ represents the total energy of the $M_2A_2Z_4$ system, and $E_M$, $E_A$, and $E_Z$ are energies of constituent atoms in their bulk form.

The catalytic performance for the HER and OER is evaluated by calculating the $\Delta G$ for each step, which is defined as:[50,51]

$$\Delta G = \Delta E + \Delta E_{ZPE} - T\Delta S + \Delta G_{pH} + \Delta G_U \quad (2)$$

where $\Delta E$ is the energy difference obtained from DFT, $\Delta E_{ZPE}$ and $\Delta S$ are the zero-point energy and entropy differences, respectively. The temperature $T$ was set to 298.15 K. $\Delta G_{pH}$ represents the free energy contribution from the proton concentration, expressed as $\Delta G_{pH} = 0.059 \times \text{pH}$. $\Delta G_U$ accounts for the influence of extra potential bias provided by the electrons or holes, calculated as $\Delta G_U = -eU$, where $U$ is the potential relative to the standard hydrogen electrode. For HER and OER, the free energy of a proton–electron pair ($H^+ + e^-$) is referenced to $1/2G_{H_2}$ under standard conditions (pH = 0, U = 0), and the free energy of gaseous $O_2$ is derived as $G_{O_2} = 2G_{H_2O} - 2\,G_{H_2} - 4.92$ eV as DFT has difficulty accurately describing the triplet ground state of $O_2$.

In photocatalytic reactions, the $U_e$ for hydrogen reduction and $U_h$ for water oxidation are related to the CBM and VBM, respectively, which can be calculated using the following equations:[40]

$$U_e = \text{CBM} - (-4.44 + 0.059 \times \text{pH}) \quad (3)$$

$$U_h = -\text{VBM} + (-4.44 + 0.059 \times \text{pH}) \quad (4)$$

## Supporting Information

The Supporting Information includes supporting tables, figures, and the energetics and structural analysis of vacancies in $M_2A_2Z_4$.

## Acknowledgements

This work was supported by the National Key Research Program of China (grant no. 2025YFE0121800, 2022YFA1503101), National Natural Science Foundation of China (grant no. 92580104, 22203058), Science and Technology Development Fund, Macau SAR (FDCT No. 0030/2022/AGJ), Collaborative Innovation Center of Suzhou Nano Science & Technology, Priority Academic Program Development of Jiangsu Higher Education Institutions (PAPD), 111 Project, and Joint International Research Laboratory of Carbon-Based Functional Materials and Devices.

**Keywords:** Two dimensional materials • $M_2A_2Z_4$ • photocatalytic overall water splitting • First-principles calculation

**Entry for the Table of Contents**

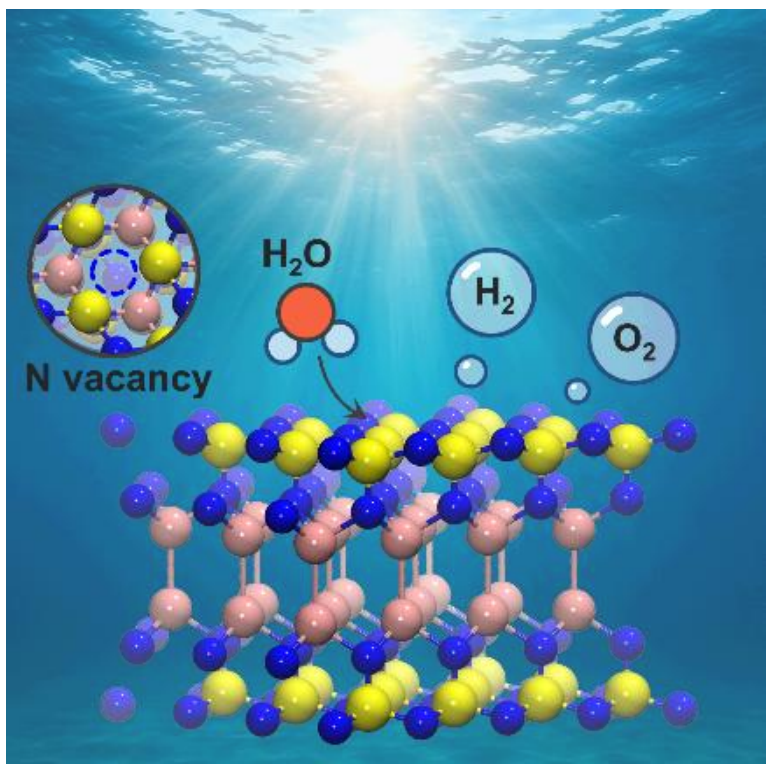

First-principles calculations are used to design and screen 108 octuple-atomic-layer two-dimensional (2D) $M_2A_2Z_4$ monolayers for photocatalytic overall water splitting. $Al_2Si_2N_4$ and $Al_2Ge_2N_4$ emerge as standout photocatalysts in both acidic and neutral conditions, while surface nitrogen vacancies markedly enhance catalytic activity for both hydrogen reduction and water oxidation reaction

## Supporting Information for

# Rational Design of Two-Dimensional Octuple-Atomic-Layer $M_2A_2Z_4$ for Photocatalytic Water Splitting

Shikai Chang[a], Dingyanyan Zhou[a], Yujin Ji*[a], Mir F. Mousavi[b], Jian Xi*[c], and Youyong Li*[a,d]

[a] State Key Laboratory of Bioinspired Interfacial Materials Science, Institute of Functional Nano & Soft Materials (FUNSOM), Soochow University, Suzhou 215123, China

[b] Department of Chemistry, Faculty of Basic Sciences, Tarbiat Modares University, Tehran

[c] Suzhou Laboratory, Suzhou 215123, China

[d] Macao Institute of Materials Science and Engineering, Macau University of Science and Technology, Taipa, Macau SAR 999078, China

Email address: yjji@suda.edu.cn (Y. Ji); xij@szlab.ac.cn (J. Xi); yyli@suda.edu.cn (Y. Li)

## Note S1. Structural properties of $M_2A_2Z_4$ candidates.

Representative POSCAR files are provided for the four configurations (*αα*, *αβ*, *βα*, and *ββ*) considered in this work. Here, *a* denotes the in-plane lattice constant, and the corresponding *a* values for all $M_2A_2Z_4$ systems are summarized in Table S2.

**1. POSCAR file of *αα* phase:**

Exampled by $Al_2Si_2N_4$

1

$\sqrt{3}a/2$ -0.5*a* 0

0 *a* 0

0 0 30

N Si Al

4 2 2

Direct

0.333 0.667 0.405

0.333 0.667 0.570

0.000 0.000 0.643

0.000 0.000 0.333

0.333 0.667 0.628

0.333 0.667 0.348

0.667 0.333 0.442

0.667 0.333 0.534

**2. POSCAR file of *αβ* phase:**

Exampled by $Al_2Si_2N_4$

1

$\sqrt{3}a/2$ -0.5*a* 0

0 *a* 0

0 0 30

N Si Al

4 2 2

Direct

0.667 0.333 0.354

0.333 0.667 0.426

0.333 0.667 0.590

0.000 0.000 0.662

0.333 0.667 0.647

0.333 0.667 0.369

0.667 0.333 0.463

0.667 0.333 0.554

**3. POSCAR file of *βα* phase:**

Exampled by $Al_2Si_2N_4$

1

$\sqrt{3}a/2$ -0.5*a* 0

0 *a* 0

0 0 30

N Si Al

4 2 2

Direct

0.333 0.667 0.342

0.667 0.333 0.414

0.000 0.000 0.578

0.667 0.333 0.650

0.667 0.333 0.357

0.000 0.000 0.635

0.333 0.667 0.542

0.333 0.667 0.451

**4. POSCAR file of *ββ* phase:**

Exampled by $Al_2Si_2N_4$

1

$\sqrt{3}a/2$ -0.5*a* 0

0 *a* 0

0 0 30

N Si Al

4 2 2

Direct

0.667 0.333 0.423

0.000 0.000 0.588

0.667 0.333 0.661

0.000 0.000 0.350

0.667 0.333 0.365

0.000 0.000 0.646

0.333 0.667 0.552

0.333 0.667 0.460

## Note S2. Computational details for the carrier mobility and STH efficiency.

The carrier mobility is obtained based on the deformation potential approximation (DPA) theory proposed by Bardeen and Shockley with the following expression[1]:

$$\mu_{2D} = \frac{e\hbar^3 C_{2D}}{k_B T m^* m_d E_1^2} \quad \text{(S1)}$$

where $k_B$ and $T$ are the Boltzmann constant and room temperature. From the equations, it can be seen that the deformation potential constant ($E_1$), the elastic modulus ($C_{2D}$), the effective mass ($m^*$) and the average effective mass are all important factors in determining the carrier mobility. They are specifically analyzed below:

The deformation potential constant ($E_1$) represents the energy change of the VBM or CBM under strain in different directions and is defined as:

$$E_1 = \Delta E/(\Delta l/l_0) \quad \text{(S2)}$$

where $\Delta E$ is the energy change under tensile and compressive strains, $l_0$ is the equilibrium lattice constant in the transport direction, and $\Delta l$ is the deformation of $l_0$.

In order to evaluate the elastic modulus ($C_{2D}$) in different directions, by the following equation:

$$(E - E_0)/S_0 = C_{2D}(\Delta l/l_0)^2/2 \quad \text{(S3)}$$

where $E$ and $E_0$ denote the total energy at strain and equilibrium, respectively, and $S_0$ denotes the lattice area at equilibrium. In the calculation process, $E_1$ and $C_{2D}$ can be obtained by linear fitting of the energy band edge positions and parabolic fitting of the total energy in different cases, respectively. In most cases, the effective mass is related to the flatness of the energy band located near the Fermi energy level. The effective mass is obtained by the following expression:

$$m^* = \hbar^2 (\frac{d^2 E_k}{dk^2})^{-1} \quad \text{(S4)}$$

where $k$ and $E_k$ denote the wave vector and the corresponding energy, respectively. The average effective mass can be calculated as follow:

$$m_d = \sqrt{m_x^* m_y^*} \tag{S5}$$

where $m_x^*$ and $m_y^*$ is the effective mass along the x and y axis, respectively.

The theoretical STH efficiency ($\eta_{\mathrm{STH}}$) is calculated as follows:

$$\eta_{\mathrm{STH}} = \frac{\Delta G \int_E^{\infty} \frac{P(\hbar\omega)}{\hbar\omega} \mathrm{d}(\hbar\omega)}{\int_0^{\infty} P(\hbar\omega)\mathrm{d}(\hbar\omega)} \tag{S6}$$

where Δ*G* is the potential difference for water splitting (1.23 eV), and *P*(*ħω*) is the AM1.5 solar energy flux at the photon energy *ħω*. *E* is the photon energy that can be utilized for water splitting redox reactions, and can be determined by:

$$E = \begin{cases} E_{\mathrm{g}}, & (\chi(\mathrm{H_2}) \geq 0.2, \chi(\mathrm{O_2}) \geq 0.6) \\ E_{\mathrm{g}} + 0.2 - \chi(\mathrm{H_2}), & (\chi(\mathrm{H_2}) < 0.2, \chi(\mathrm{O_2}) \geq 0.6) \\ E_{\mathrm{g}} + 0.6 - \chi(\mathrm{O_2}), & (\chi(\mathrm{H_2}) \geq 0.2, \chi(\mathrm{O_2}) < 0.6) \\ E_{\mathrm{g}} + 0.8 - \chi(\mathrm{H_2}) - \chi(\mathrm{O_2}), & (\chi(\mathrm{H_2}) < 0.2, \chi(\mathrm{O_2}) < 0.6) \end{cases} \tag{S7}$$

where $\chi(H_2)$ is the difference between the conduction band minimum (CBM) of the candidate and the hydrogen reduction potential, and $\chi(O_2)$ is the difference between the valence band maximum (VBM) of the candidate and the water oxidation potential.

## Note S3. Energetics and Structural Analysis of Vacancies in $M_2A_2Z_4$

The formation energy of a single A or Z vacancy at the surface of a pristine $M_2A_2Z_4$ supercell is calculated as:

$$E_f^v = E_v - E_{M_2A_2Z_4} + nE_{atom} \quad (S8)$$

where $E_v$, $E_{M_2A_2Z_4}$ and $E_{atom}$ represent the total energy of the supercell containing the vacancy, the energy of the pristine $M_2A_2Z_4$ supercell, and the energy of the isolated atom corresponding to the vacancy, respectively; $n$ is the number of vacancies.

For $Al_2Si_2N_4$ and $Al_2Ge_2N_4$, the formation energies of a single N and Si vacancy in $Al_2Si_2N_4$ are 4.96 eV and 35.30 eV, respectively, and in $Al_2Ge_2N_4$, the corresponding values of N and Ge vacancy are 2.30 eV and 5.49 eV, respectively. All calculated formation energy values are positive, indicating that energy input is required to form these defects, and vacancies can stably exist after formation. Furthermore, the formation energy of the N vacancy is significantly lower than that of the Si/Ge vacancy, implying that N vacancy is favor to form compared to Si/Ge vacancy on the surface of $Al_2Si_2N_4$ and $Al_2Ge_2N_4$. Structurally, the introduction of N vacancies leads to only slight distortion in both $Al_2Si_2N_4$ and $Al_2Ge_2N_4$, whereas Si/Ge vacancies introduce noticeable disruption of the monolayers.

**Table S1.** Formation energies of different phases of $M_2A_2Z_4$, units in eV/atom.

| | *αα* | *αβ* | *βα* | *ββ* |
|---|---|---|---|---|
| $Al_2Si_2N_4$ | **-0.972** | -0.970 | -0.963 | -0.962 |
| $Ga_2Si_2N_4$ | **-0.660** | -0.656 | -0.653 | -0.656 |
| $In_2Si_2N_4$ | **-0.270** | -0.269 | -0.269 | -0.270 |
| $Al_2Ge_2N_4$ | -0.411 | **-0.412** | -0.407 | -0.404 |
| $Ga_2Ge_2N_4$ | **-0.132** | -0.131 | -0.129 | -0.129 |
| $In_2Ge_2N_4$ | **0.163** | 0.165 | 0.165 | 0.166 |
| $Al_2Sn_2N_4$ | **-0.118** | -0.118 | -0.118 | -0.111 |
| $Ga_2Sn_2N_4$ | 0.107 | **0.106** | 0.112 | 0.114 |
| $In_2Sn_2N_4$ | 0.260 | **0.255** | 0.258 | 0.257 |
| $Al_2Si_2P_4$ | -0.238 | **-0.239** | -0.238 | -0.236 |
| $Ga_2Si_2P_4$ | **-0.207** | -0.205 | -0.206 | -0.205 |
| $In_2Si_2P_4$ | -0.029 | **-0.029** | -0.029 | -0.028 |
| $Al_2Ge_2P_4$ | -0.163 | -0.165 | **-0.165** | -0.162 |
| $Ga_2Ge_2P_4$ | -0.127 | -0.129 | -0.129 | **-0.134** |
| $In_2Ge_2P_4$ | **0.022** | 0.022 | 0.022 | 0.024 |
| $Al_2Sn_2P_4$ | -0.148 | -0.146 | **-0.152** | -0.144 |
| $Ga_2Sn_2P_4$ | -0.114 | -0.113 | **-0.114** | -0.110 |
| $In_2Sn_2P_4$ | -0.026 | **-0.028** | -0.028 | -0.027 |
| $Al_2Si_2As_4$ | -0.141 | -0.143 | **-0.144** | -0.139 |
| $Ga_2Si_2As_4$ | -0.135 | -0.131 | -0.132 | **-0.135** |
| $In_2Si_2As_4$ | 0.001 | -0.006 | **-0.007** | -0.004 |
| $Al_2Ge_2As_4$ | -0.132 | -0.136 | **-0.137** | -0.132 |
| $Ga_2Ge_2As_4$ | -0.117 | -0.129 | **-0.131** | -0.128 |
| $In_2Ge_2As_4$ | -0.011 | -0.021 | **-0.022** | -0.019 |
| $Al_2Sn_2As_4$ | -0.159 | -0.162 | **-0.165** | -0.160 |
| $Ga_2Sn_2As_4$ | -0.153 | -0.156 | **-0.159** | -0.155 |
| $In_2Sn_2As_4$ | -0.096 | **-0.099** | -0.097 | -0.098 |

**Table S2.** Phase, optimized lattice parameters (*a*), and PBE and HSE band gaps of $M_2A_2Z_4$.

| | phase | *a* (Å) | $E_g^{PBE}$ (eV) | $E_g^{HSE06}$ (eV) |
|---|---|---|---|---|
| $Al_2Si_2N_4$ | *αα* | 2.950 | 1.21 | 1.77 |
| $Ga_2Si_2N_4$ | *αα* | 2.993 | 2.85 | 3.69 |
| $In_2Si_2N_4$ | *αα* | 3.117 | 1.48 | 2.54 |
| $Al_2Ge_2N_4$ | *αβ* | 3.066 | 0.82 | 1.76 |
| $Ga_2Ge_2N_4$ | *αα* | 3.110 | 1.64 | 2.77 |
| $In_2Ge_2N_4$ | *αα* | 3.244 | 0.62 | 1.51 |
| $Al_2Sn_2N_4$ | *αα* | 3.273 | 0.07 | 0.80 |
| $Ga_2Sn_2N_4$ | *αβ* | 3.320 | 0.35 | 1.24 |
| $In_2Sn_2N_4$ | *αβ* | 3.460 | 0.10 | 0.68 |
| $Al_2Si_2P_4$ | *αβ* | 3.647 | 1.20 | 1.94 |
| $Ga_2Si_2P_4$ | *αα* | 3.660 | 1.11 | 1.79 |
| $In_2Si_2P_4$ | *αβ* | 3.776 | 0.55 | 1.29 |
| $Al_2Ge_2P_4$ | *βα* | 3.724 | 1.01 | 1.66 |
| $Ga_2Ge_2P_4$ | *ββ* | 3.734 | 1.03 | 1.70 |
| $In_2Ge_2P_4$ | *αα* | 3.866 | 0.18 | 0.86 |
| $Al_2Sn_2P_4$ | *βα* | 3.879 | 0.98 | 1.58 |
| $Ga_2Sn_2P_4$ | *βα* | 3.879 | 1.06 | 1.66 |
| $In_2Sn_2P_4$ | *βα* | 3.776 | – | – |
| $Al_2Si_2As_4$ | *βα* | 3.821 | 0.98 | 1.67 |
| $Ga_2Si_2As_4$ | *ββ* | 3.832 | 0.72 | 1.38 |
| $In_2Si_2As_4$ | *βα* | 3.943 | 0.05 | 0.55 |
| $Al_2Ge_2As_4$ | *βα* | 3.894 | 0.58 | 1.19 |
| $Ga_2Ge_2As_4$ | *βα* | 3.898 | 0.42 | 0.97 |
| $In_2Ge_2As_4$ | *βα* | 4.034 | 0.03 | 0.26 |
| $Al_2Sn_2As_4$ | *βα* | 4.028 | 0.73 | 1.27 |
| $Ga_2Sn_2As_4$ | *βα* | 4.037 | 0.72 | 1.27 |
| $In_2Sn_2As_4$ | *αβ* | 4.226 | 0.03 | 0.98 |

**Table S3.** Calculated the effective mass $m^{*}$, deformation potential constant $E_1$, elastic modulus $C_{2D}$, carrier mobility $\mu$ along the x and y direction at 300 K.

| Material | | $m^{*}$ $(m_0)$ | | $E_1$ (eV) | | $C_{2D}$ (N/m) | $\mu$ ($cm^2V^{-1}s^{-1}$) | |
|---|---|---|---|---|---|---|---|---|
| | | e | h | e | h | e = h | e | h |
| $Al_2Si_2N_4$ | x | 0.88 | 1.18 | 3.33 | 8.52 | 563.82 | 2504.04 | 207.36 |
| | y | 0.27 | 0.38 | 3.33 | 5.36 | 568.98 | 8235.10 | 1643.41 |
| $Al_2Ge_2N_4$ | x | 0.27 | 1.41 | 11.39 | 7.71 | 481.86 | 1099.94 | 149.64 |
| | y | 0.27 | 0.46 | 14.37 | 7.89 | 480.55 | 686.73 | 432.56 |

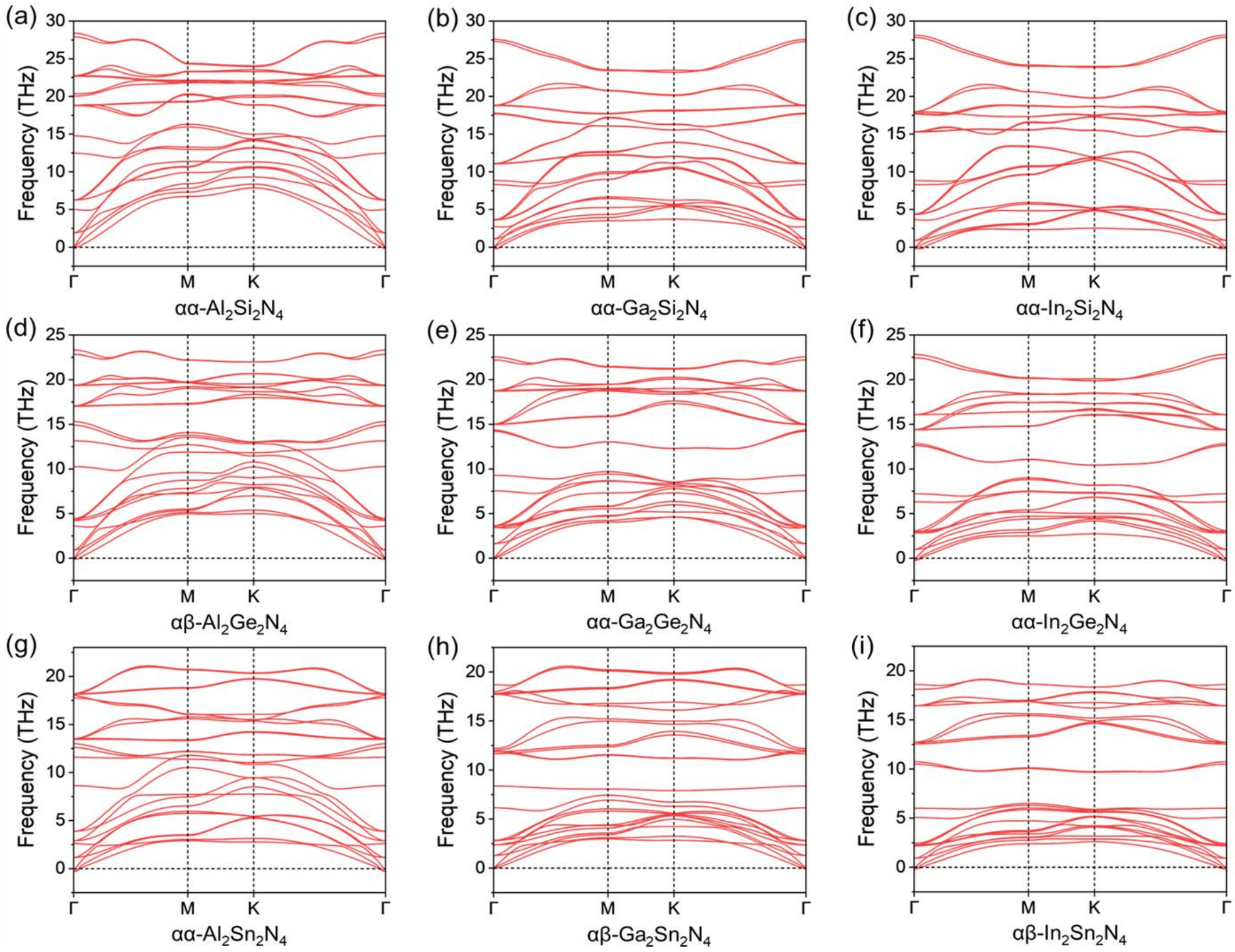

**Figure S1**. Phonon spectrum of $M_2A_2N_4$ monolayers.

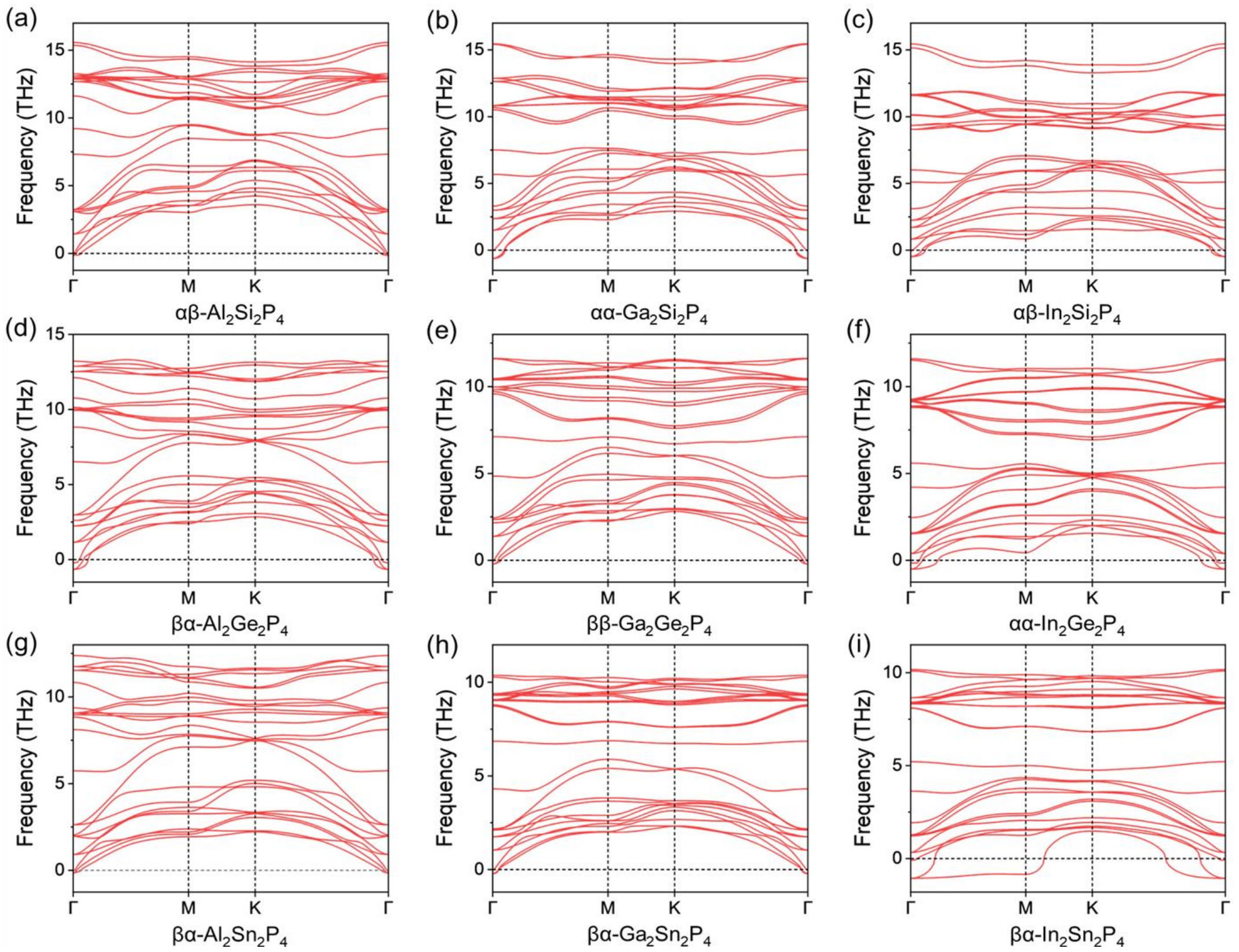

**Figure S2**. Phonon spectrum of $M_2A_2P_4$ monolayers.

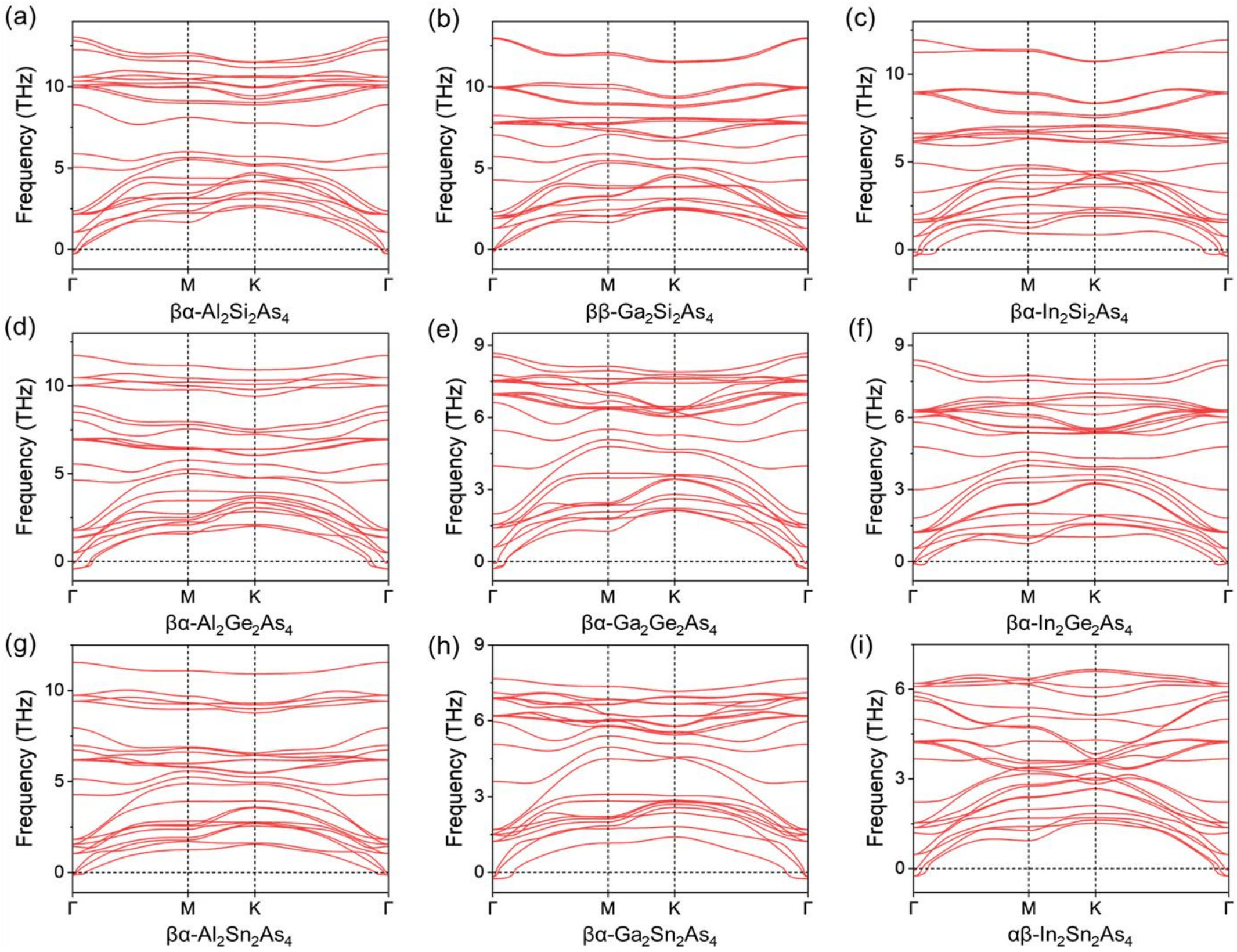

**Figure S3**. Phonon spectrum of $M_2A_2As_4$ monolayers.

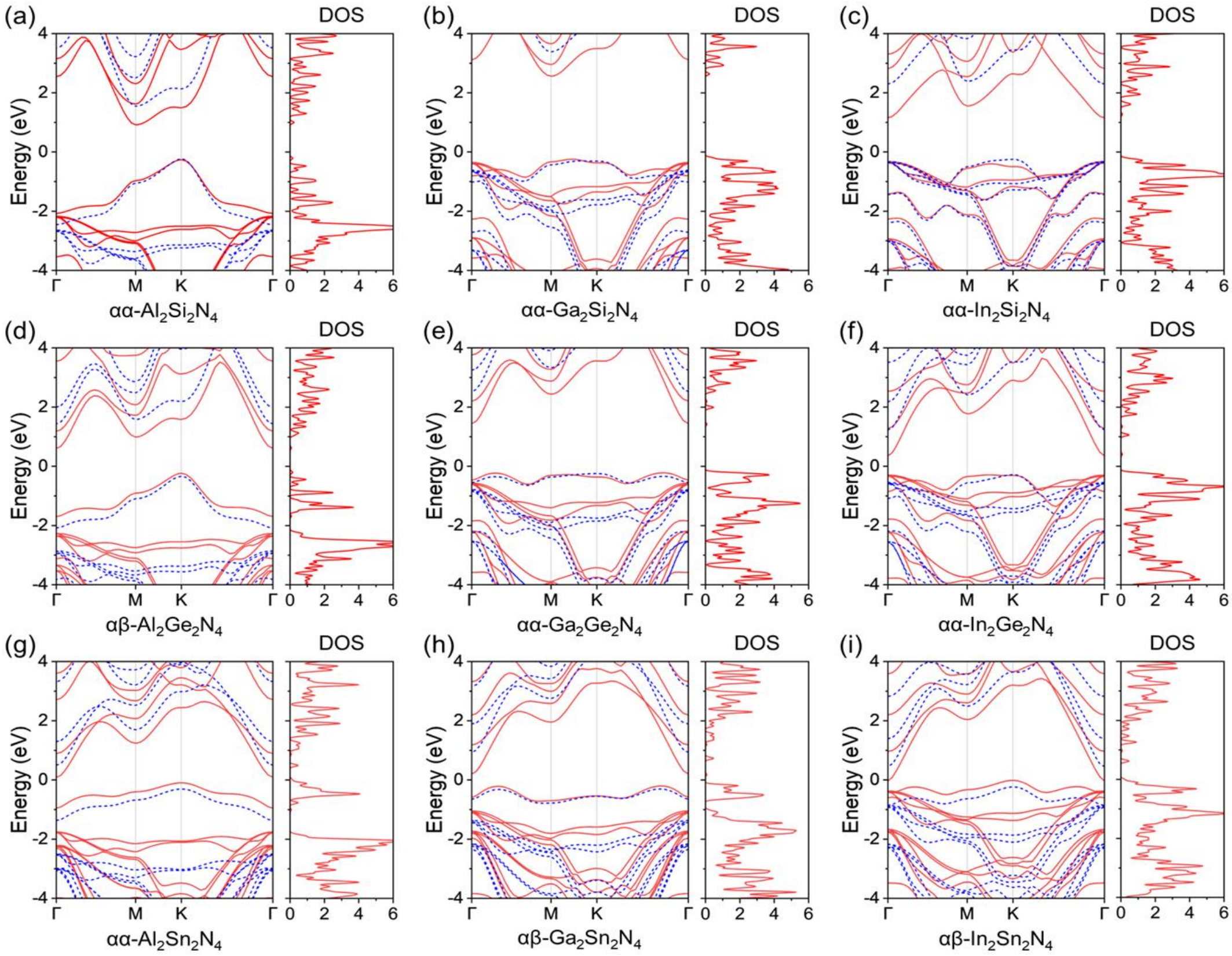

**Figure S4.** Electronic band structures of $M_2A_2N_4$ monolayers. The PBE and HSE bands are shown in red solid and blue dash lines, respectively.

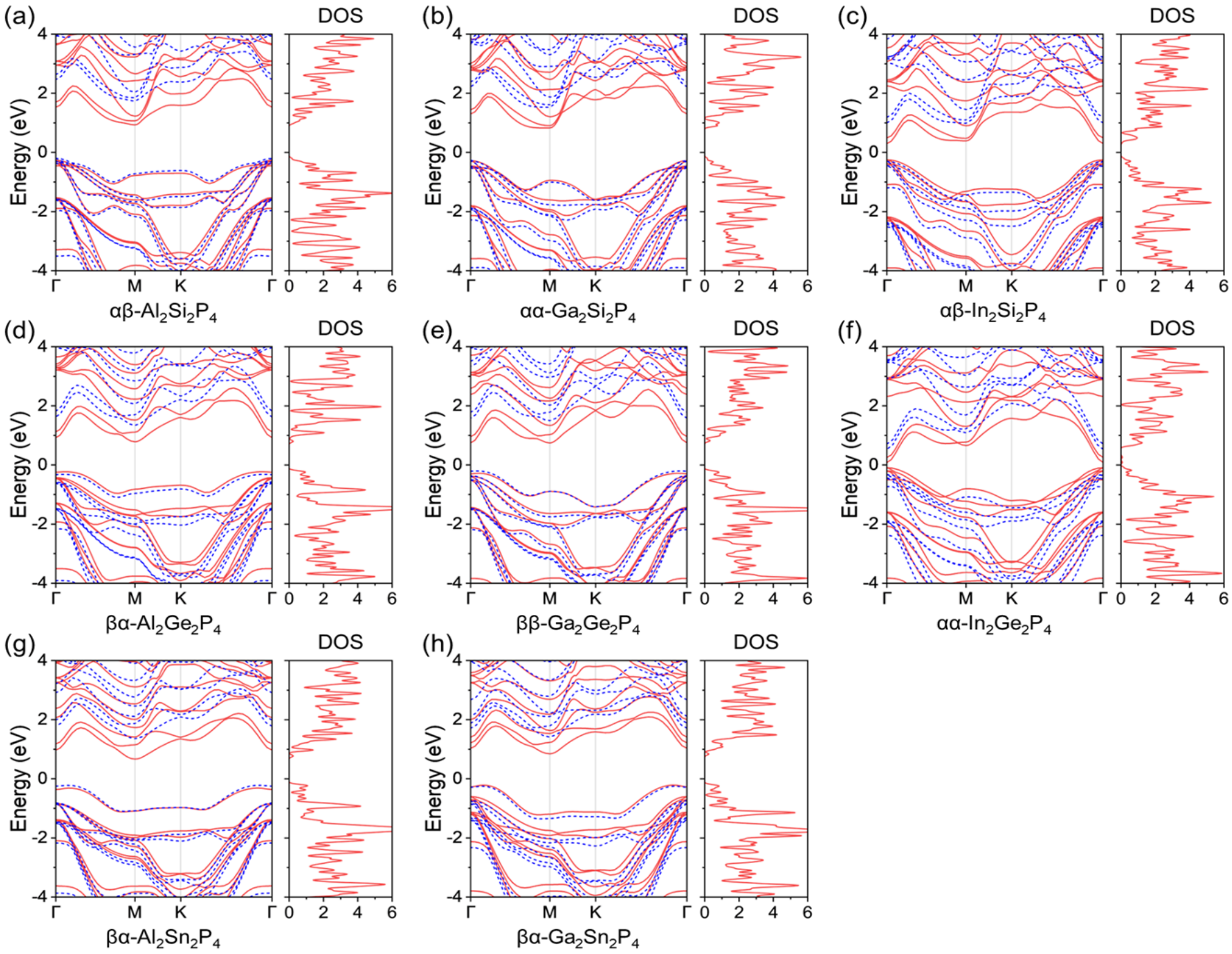

**Figure S5.** Electronic band structures of $M_2A_2P_4$ monolayers. The PBE and HSE bands are shown in red solid and blue dash lines, respectively.

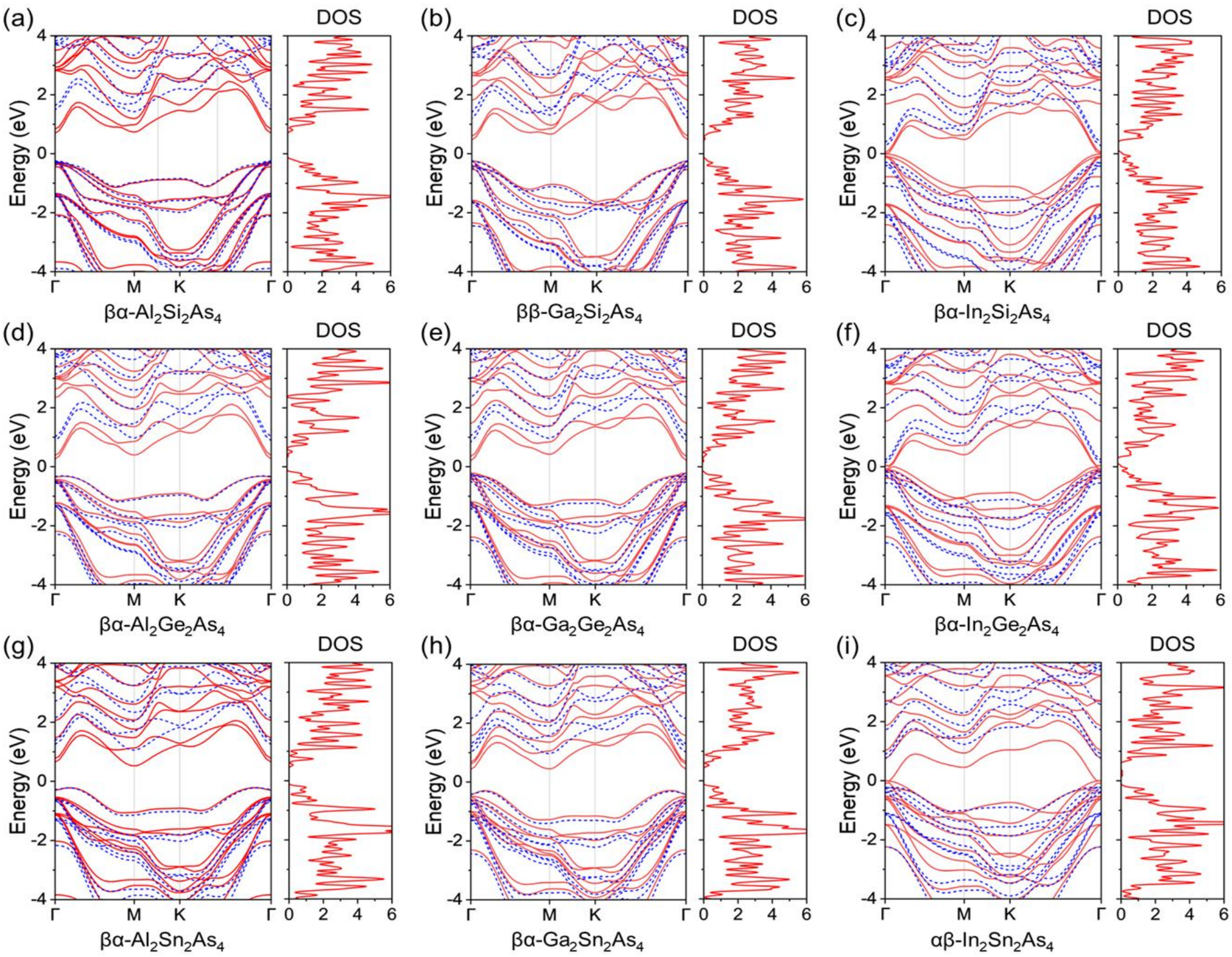

**Figure S6.** Electronic band structures of $M_2A_2As_4$ monolayers. The PBE and HSE bands are shown in red solid and blue dash lines, respectively.

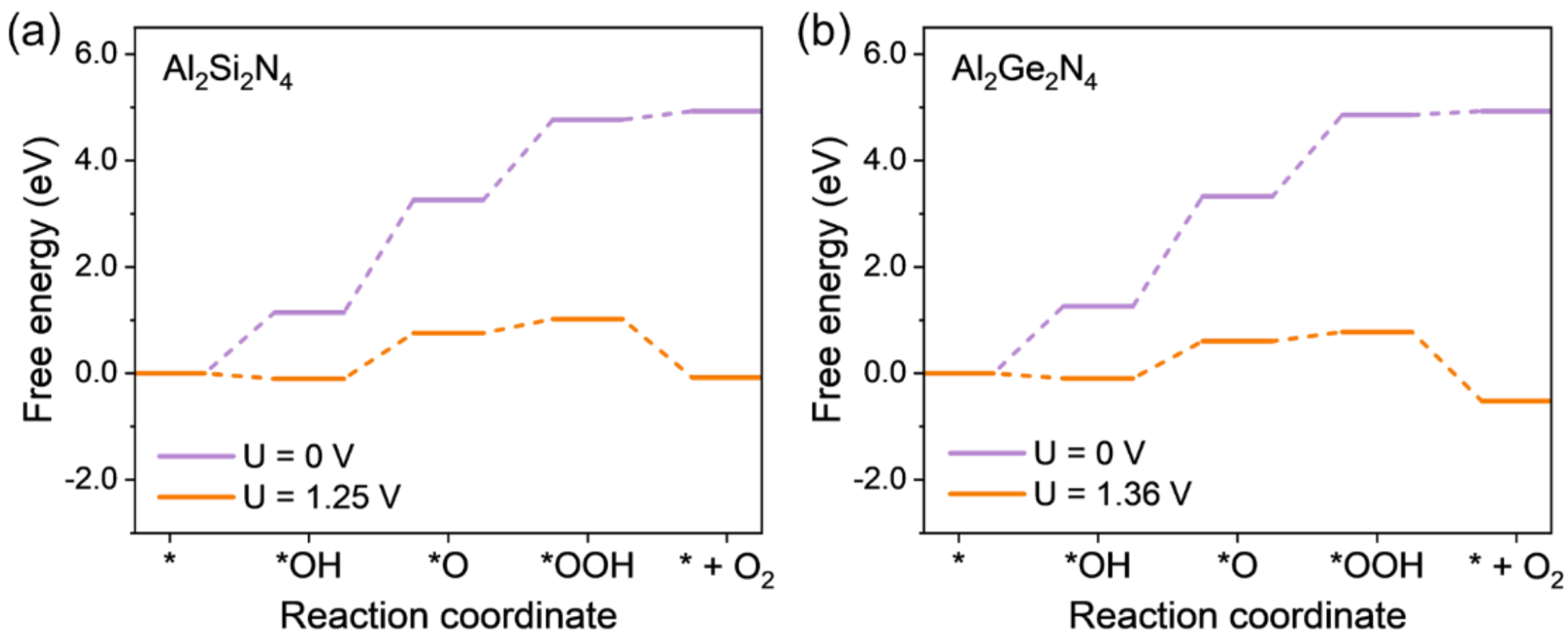

**Figure S7.** Gibbs free energy profiles of the OER on pristine (a) $Al_2Si_2N_4$ and (b) $Al_2Ge_2N_4$. The purple line represents conditions in dark at pH = 0, the orange line corresponds to conditions of light irradiation at pH = 0.

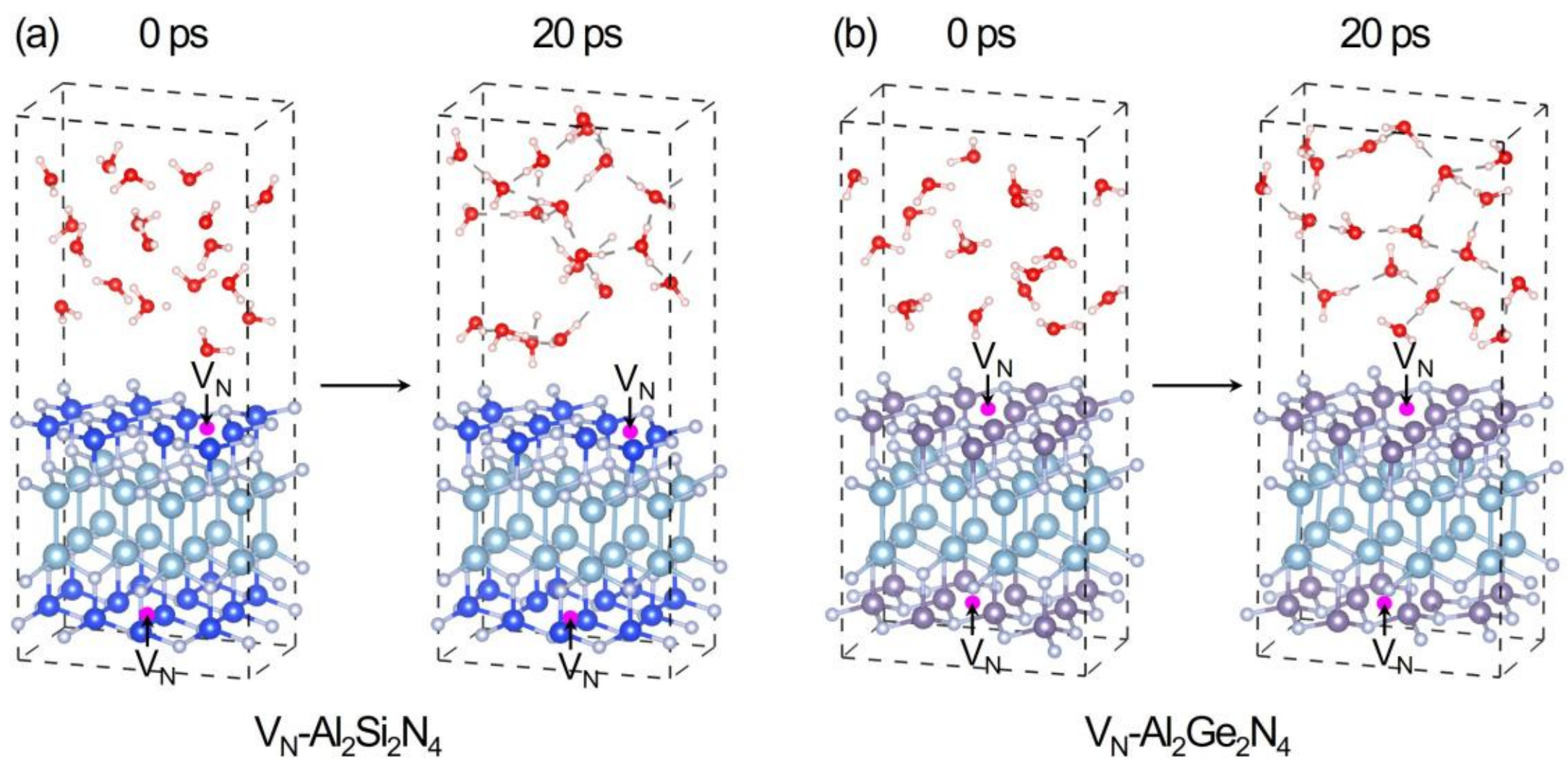

**Figure S8.** Atomic structures of (a) $Al_2Si_2N_4$ and (b) $Al_2Ge_2N_4$ with N vacancy defects.